\documentclass[twocolumn,twoside,slac]{revtex4}
\usepackage{graphicx}
\usepackage{fancyhdr}
\begin{document}

\title{TPC tracking and particle identification in high-density environment }

\author{Y.Belikov, M.Ivanov, K.Safarik }
\affiliation{CERN, Switzerland}
\author{J.Bracinik}
\affiliation{Comenius University, Bratislava}

\begin{abstract}
Track finding and fitting algorithm in the ALICE Time projection
chamber (TPC) based on Kalman-filtering is presented.
Implementation of particle identification (PID) using d$E$/d$x$
measurement is discussed. Filtering and PID algorithm is able to
cope with non-Gaussian noise as well as with ambiguous
measurements in a high-density environment. The occupancy can
reach up to 40\% and due to the overlaps, often the points along
the track are lost and others are significantly displaced. In the
present algorithm, first, clusters are found and the space points
are reconstructed. The shape of a cluster provides information
about overlap factor. Fast spline unfolding algorithm is applied
for points with distorted shapes. Then, the expected space point
error is estimated using information about the cluster shape and
track parameters. Furthermore, available information about local
track overlap is used. Tests are performed on simulation data sets
to validate the analysis and to gain practical experience with the
algorithm.
\end{abstract}

\maketitle \thispagestyle{fancy}


\section{Introduction}

Track finding for the predicted particle densities is one of the
most challenging tasks in the ALICE experiment \cite{ALICE}. It is
still under development and here the current status is reported.
Track finding is based on the Kalman-filtering approach.
Kalman-like algorithms are widely used in high-energy physics
experiments and their advantages and shortcomings are well known.

There are two main disadvantages of the Kalman filter, which
affect the tracking in the ALICE TPC \cite{TPCTDR}. The first is
that before applying the Kalman-filter procedure, clusters have to
be reconstructed.  Occupancies up to 40\%  in the inner sectors of
the TPC and up to 20\% in the outer sectors are expected; clusters
from different tracks may be overlapped; therefore a certain
number of the clusters are lost, and the others may be
significantly displaced. These displacements are rather hard to
take into account. Moreover, these displacements are strongly
correlated depending on the  distance between two tracks.

The other disadvantage of the Kalman-filter tracking is that it
relies essentially on the determination of good `seeds' to start a
stable filtering procedure. Unfortunately, for the tracking in the
ALICE TPC the seeds using the TPC data themselves have to be
constructed. The TPC is a key starting point for the tracking in
the entire ALICE set-up.  Until now, practically none of the other
detectors have been able to provide the initial information about
tracks.

 On the other hand, there is a whole list of very attractive properties of
the Kalman-filter approach.
\begin{itemize}

\item It is a method for simultaneous track recognition and
fitting.

\item There is a possibility to reject incorrect space points `on
the fly', during the only tracking pass. Such incorrect points can
appear as a consequence of the imperfection of the cluster finder.
They may be due to noise or they may be points from other tracks
accidentally captured in the list of points to be associated with
the track under consideration. In the other tracking methods one
usually needs an additional fitting pass to get rid of incorrectly
assigned points.

\item In the case of substantial multiple scattering, track
measurements are correlated and therefore large matrices (of the
size of the number of measured points) need to be inverted during
a global fit. In the Kalman-filter procedure we only have to
manipulate up to $5 \times 5$ matrices (although many times, equal
to the number of measured points), which is much faster.

\item Using this approach one can handle multiple scattering and
energy losses in a simpler way than in the case of global methods.

\item Kalman filtering is a natural way to find the extrapolation
of a track from one detector to another (for example from the TPC
to the ITS or to the TRD).
\end{itemize}

The following parametrization for the track was chosen:
\begin{eqnarray}
y(x)= y_0 - \frac{\displaystyle 1}{\displaystyle C}
           \sqrt{1-({ C}x-{\eta})^2}  \\
z(x)= z_0-\frac{\displaystyle\tan\lambda}{\displaystyle C}
           \arcsin({ C}x-{\eta})
\end{eqnarray}
The  state vector $x^{\rm{T}}$ is given by the local track
position {\it{x}}, {\it{y}} and {\it{z}}, by a curvature {\it{C}},
local {\it{x}}$_0$ position of the helix center, and dip angle
$\lambda$:
\begin{eqnarray}
       {\bf x}^{\rm{T}}=(y,\ z,\ C,\ \tan\lambda,\ \eta), ~\eta\equiv Cx_0
\end{eqnarray}

Because of high occupancy the standard Kalman filter approach was
modified. We tried to find maximum additional possible information
which can be used during cluster finding, tracking and particle
identification. Because of too many degrees of freedom (up to 220
million 10-bit samples) we have to find a smaller number of
orthogonal parameters.

To enable using the optimal combination of local and global
information about the tracks and clusters, the parallel Kalman
filter tracking method was proposed. Several hypothesis are
investigated in parallel. The global tracking approach such as
Hough transform was considered only for seeding of track
candidates. In the following, the additional information which was
used will be underlined.

\section{Accuracy of local coordinate measurement}

\begin{figure}[t]
%\centering
\includegraphics[width=60mm,angle=-90]{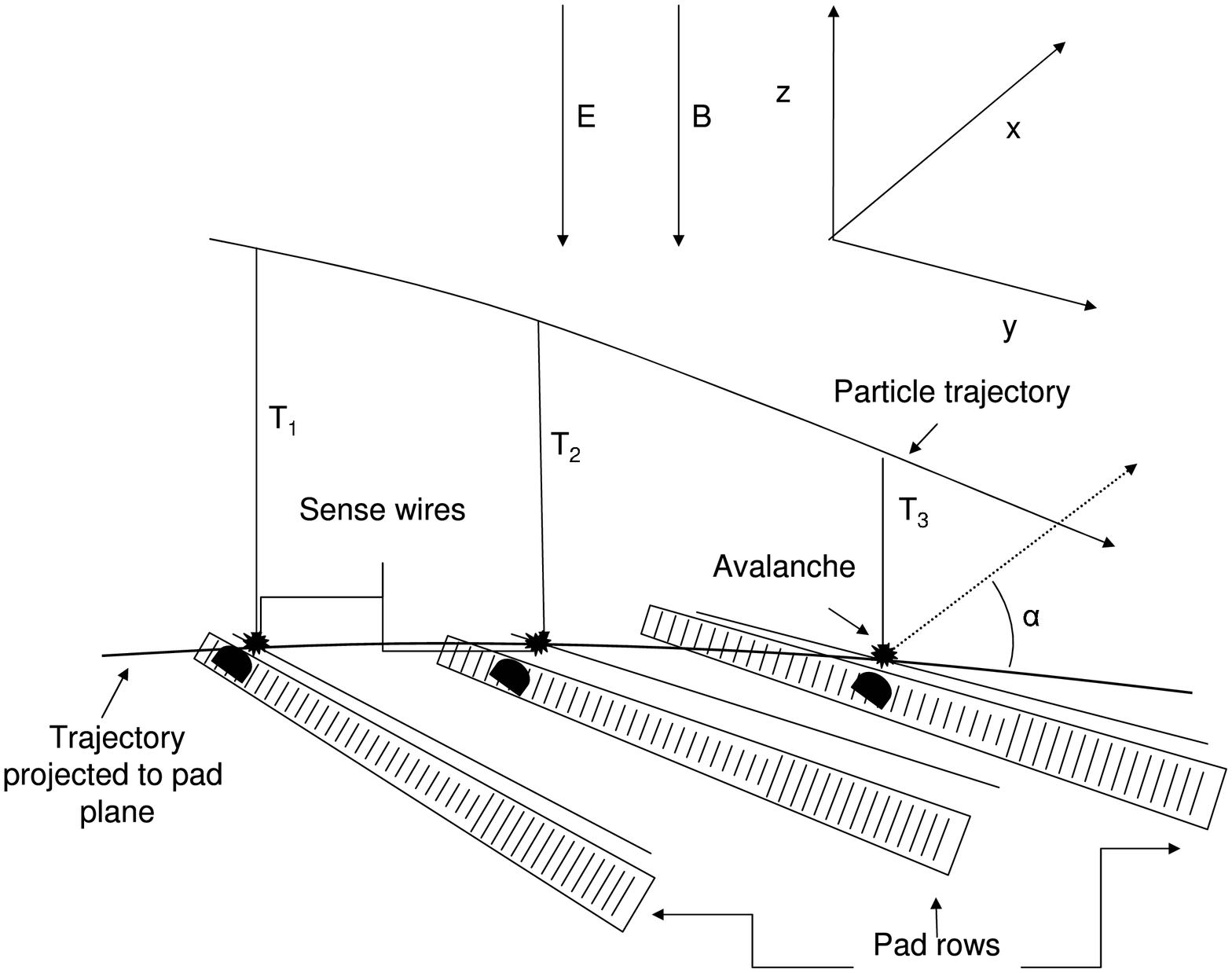}
\includegraphics[width=60mm,angle=-90]{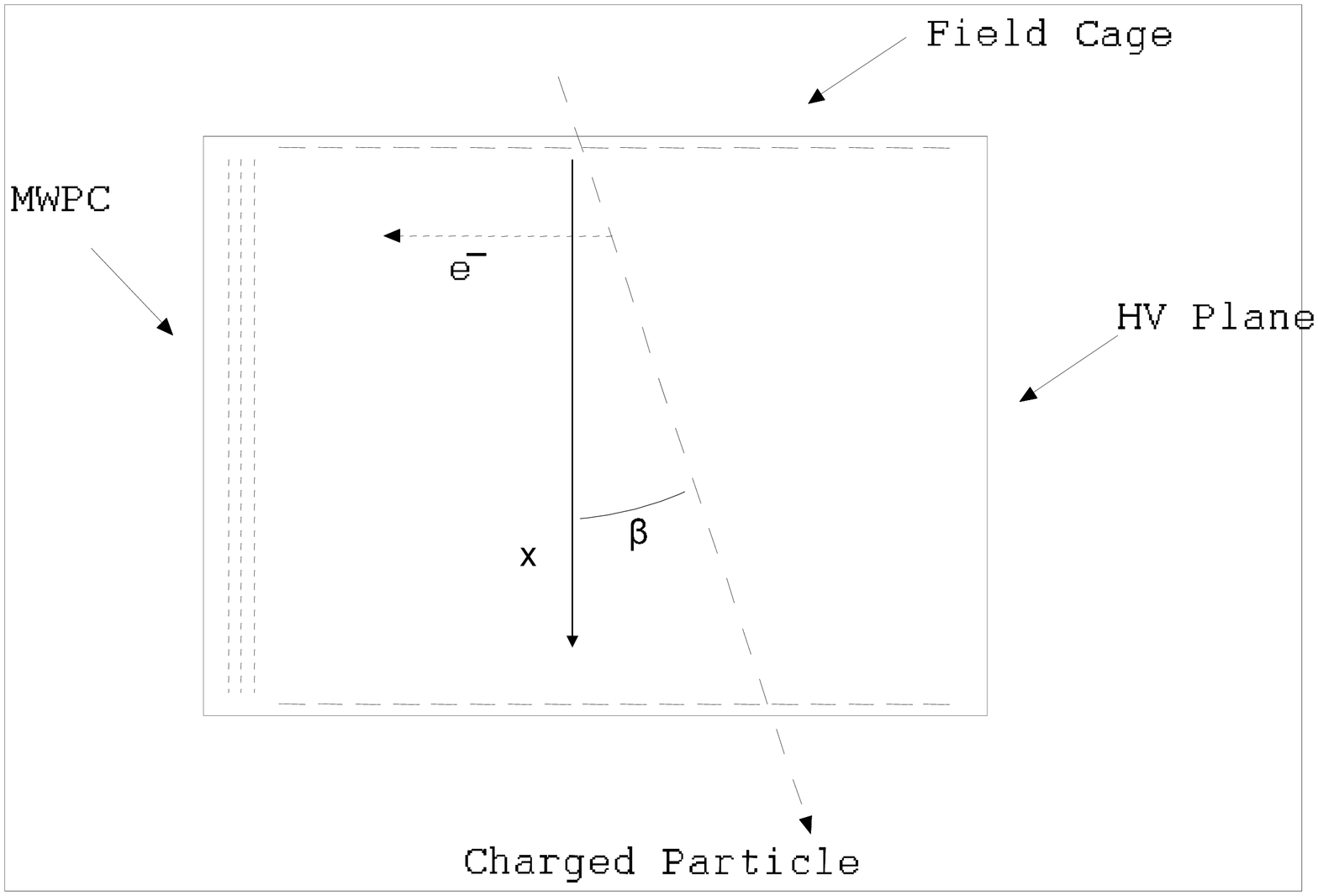}
\caption{Schematic view of the detection process in TPC  (upper
part - perspective view, lower part - side view).} \label{figTPC}
\end{figure}

The accuracy of the coordinate measurement is limited by a track
angle which spreads ionization and by diffusion which amplifies
this spread.

The track direction with respect to pad plane is given by two
angles $\alpha$ and $\beta$ (see fig.~\ref{figTPC}). For the
measurement along the pad-row, the angle $\alpha$ between the
track projected onto the pad plane and pad-row is relevant. For
the measurement of the the drift coordinate ({\it{z}}--direction)
it is the angle $\beta$ between the track and {\it{z}} axis
(fig.~\ref{figTPC}).

The ionization electrons are randomly distributed along the
particle trajectory. Fixing the reference {\it{x}} position of an
electron at the middle of pad-row, the {\it{y}} (resp. {\it{z}})
position of the electron is a random variable characterized by
uniform distribution with the width $L_{\rm{a}}$, where
$L_{\rm{a}}$ is given by the pad length $L_{\rm{pad}}$ and the
angle $\alpha$ (resp. $\beta$):
\[L_{\rm{a}}=L_{\rm{pad}}\tan\alpha\]

The diffusion smears out the position of the electron with
gaussian probability distribution with $\sigma_{\rm{D}}$.
Contribution of the $\mathbf{E{\times}B}$ and unisochronity
effects for the  Alice TPC are negligible. The typical resolution
in the case of ALICE TPC is on the level of
$\sigma_{y}\sim$~0.8~mm and $\sigma_{z}\sim$~1.0~mm integrating
over all clusters in the TPC.

\subsection{Gas gain fluctuation effect}

Being collected on sense wire, electron is "multiplied" in strong
electric field. This multiplication is subject of a large
fluctuations, contributing to the cluster position resolution.
Because of these fluctuations the center of gravity of the
electron cloud can be shifted.

Each electron is amplified independently. However, in the
reconstruction electrons are not treated separately. The Centre Of
Gravity  (COG) of the cluster is usually used as an estimation for
the local track position. The influence of the gas gain
fluctuation to the reconstructed point characteristic can be
described by a simple model, introducing a weighted COG
$X_{\rm{COG}}$
\begin{eqnarray}
    X_{\rm{COG}}=\frac{\sum_{i=1}^{N}{g_ix_i}}{\sum_{i=1}^N{g_i}},
\label{eqCOGdefGG}
\end{eqnarray}
where {\it{N}} is the total number of electrons in the cluster and
$g_i$ is a random variable equal to a gas amplification for given
electron.

The mean value of $X_{\rm{COG}}$ is equal to the mean value
$\overline{x}$ of the original distribution of electrons
\begin{eqnarray}
      \overline{X_{COG}}=
    \overline{\frac{\sum_{i=1}^{N}{g_ix_i}}{\sum_{i=1}^N{g_i}}}
    =\overline{x}\overline{\frac{\sum_{i=1}^{N}{g_i}}
    {\sum_{i=1}^N{g_i}}} =\overline{x}.
\label{eqCOGMeanGG}
\end{eqnarray}

However, the same is not true for the dispersion of the position,
%$\sigma^2_{X_{COG}}\sigma_x^2$:
%\begin {center}
\begin{eqnarray}
    \lefteqn{ \sigma^2_{X_{\rm{COG}}}
    =\overline{X_{\rm{COG}}^2}-\overline{X_{\rm{COG}}}^2=}\nonumber\\&&{}
    =\overline{\left(\frac{1}{\sum_{i=1}^N{g_i}}\sum_{i=1}^{N}{g_ix_i}
    \right)^2}-\overline{x}^2=
    \nonumber\\
    &&{}=\overline{\frac{{\sum\sum{x_ix_jg_ig_j}}}{{\sum\sum{g_ig_j}}}}-
    \overline{x}^2=
    \nonumber\\&&{}=
    \overline{x^2}\overline{\frac{\sum_i{g_i^2}}{\sum\sum{g_ig_j}}}-
    \overline{x}^2
    \overline{\frac{\sum\sum{g_ig_j}-\sum\sum_{i\ne{j}}{g_ig_j}}
    {\sum\sum{g_ig_j}}}= \nonumber\\&&
    =\left(\overline{x^2}-\overline{x}^2\right)
    \overline{\frac{\sum{g_i^2}}{\sum\sum{g_ig_j}}}=
    \sigma_x^2\overline{\frac{\sum{g_i^2}}{\sum\sum{g_ig_j}}}=
    \nonumber\\
    &&{}=\frac{\sigma_x^2}{N}{\times}G_{\rm{gfactor}}^2
\label{eqCOGSigmaGG}
\end{eqnarray}

where
\begin{eqnarray}
      G_{\rm{gfactor}}^2 = N\overline{\frac{\sum{g_i^2}}{\sum\sum{g_ig_j}}}
\label{eqCOGGGfactor0}
\end{eqnarray}

The diffusion term is effectively multiplied by gas gain factor
$G_{\rm{gfactor}}$. For sufficiently large number of electrons,
when $g_i^2$ and $\sum\sum{g_ig_j}$ are quasi independent
variables, equation (\ref{eqCOGGGfactor0}) can be transformed to
the following

\begin{eqnarray}
    \lefteqn{G_{\rm{gfactor}}^2 \approx
    N\frac{\overline{\sum{g_i^2}}}
    {\overline{\sum\sum{g_ig_j}}}}\nonumber\\
    &&{} =
    N\frac{N\overline{g^2}}{N(N-1)\overline{g}^2+N\overline{g^2}}=
    \nonumber\\
    &&{} =N\frac{ \left(\sigma_g^2/\overline{g}^2+1 \right)}
    {N+\sigma_g^2/\overline{g}^2}
\label{eqCOGGGfactorE}
\end{eqnarray}

Gas gain fluctuation of the gas detector working in proportional
regime is described with the exponential distribution with the
mean value $\bar{g}$ and r.m.s.
\begin{eqnarray}
        \sigma_{\rm{g}} =\bar{g}
\label{eqSigmaexp}
\end{eqnarray}

Substituting $\sigma_{\rm{g}}$ into equation
(\ref{eqCOGGGfactorE})
\begin{eqnarray}
    G_{\rm{gfactor}}^2 =\frac{2N}{N+1}.
\label{eqCOGGGfactorR}
\end{eqnarray}

Gas multiplication fluctuation in chamber  deteriorates
$\sigma_{X_{\rm{COG}}}$  by a factor of about ${\sqrt{2}}$. The
prediction of this model is in good agreement with results from
the simulation.

\subsection{Secondary ionization effect}

Charged particle penetrating the gas of the detector produces
{\it{N}} primary electrons. Primary electron {\it{i}} produces
$n_{\rm{s}}^i-1$ secondary electrons. Each of these electrons is
amplified in the electric field by a factor of $g_j$.

Each primary cluster is characterized by a position $x_i$ with
mean value $\overline{x}$ and $\sigma_x$. The COG given by
equation (\ref{eqCOGdefGG}) is modified to the following form:

\begin{eqnarray}
    X_{\rm{COG}}=\frac{1}{\sum_{i=1}^N\sum_{j=1}^{n_i}{g_j^{i}}}
    \sum_{i=1}^{N}{x_i}\sum_{j=1}^{n_i}{g_j^{i}}.
\label{eqCOGdefGGPIO}
\end{eqnarray}
A new variable $G_n$ is introduced as the total electron gain:
\begin{eqnarray}
    G_n=\sum_{j=1}^{n}{g_j}.
\label{eqGNdef}
\end{eqnarray}

Knowing the distribution of {\it{n}} and {\it{g}} and assuming
that {\it{n}} and {\it{g}} are independent variables  the mean
value and variance of the $G_n$ can be expressed as:

\begin{eqnarray}
    \lefteqn{
    \overline{G_n}=\overline{n}\overline{g}} \\
    &&{}
    \frac{\sigma^2_{G_n}}{\overline{G_n^2}}=
    \frac{\sigma^2_n}{\overline{n}^2}+
    \frac{\sigma^2_g}{\overline{g}^2}
    \frac{1}{\overline{n}}
\label{eqGNsigma}
\end{eqnarray}

Inserting $G_n$ into equation (\ref{eqCOGdefGGPIO}) results in an
equation similar to the equation (\ref{eqCOGdefGG}).

Multiplicative factor $G_{\rm{Lfactor}}$ is defined as an analog
of $G_{\rm{gfactor}}$, from the equation (\ref{eqCOGGGfactor0})
\begin{eqnarray}
    G_{\rm{Lfactor}}^2 =  N\frac{\overline{\sum{G_i^2}}}
    {\overline{\sum\sum{G_iG_j}}}.
\label{eqCOGLfactor0}
\end{eqnarray}

Using the new variable $G_n$ and simply replacing  gas gain
{\it{g}} by $G_n$ in the similar way as in equation
(\ref{eqCOGGGfactorE}) does not work. For $1/E^{2}$
parametrization of secondary ionization process
$\sigma^2_{G_n}/\overline{G_n}$ goes to infinity and thus
$\sigma^2_{X_{COG}}=\sigma_x^2$. Moreover $G_i^2$ and
$\sum\sum{G_iG_j}$ are not quasi independent as the sum
$\sum\sum{G_iG_j}$ could be given by one "exotic" electron
cluster. Approximations used for deriving the equation
(\ref{eqCOGGGfactorE}) are not valid for secondary ionization
effect.

In order to estimate the impact of this effect on COG  equation
(\ref{eqCOGLfactor0}) has to be solved numerically. Simulation
showed that $G_{\rm{Lfactor}}$ does not depend strongly on the cut
used for maximum number of electrons created in the process of
secondary ionization. A change of the cut,  from 1000 electrons up
produces a change of about 3\% in $G_{\rm{Lfactor}}$.

Equation (\ref{eqCOGGGfactorE}) is not applicable in this
situation because of the infinity of the $\sigma_G$. According to
the simulation, the threshold  on the number of electrons in the
cluster  has a little influence to the resulting
$G_{\rm{Lfactor}}$. Therefore we fit simulated $G_{\rm{Lfactor}}$
with formula (\ref{eqCOGGGfactorE}) where
$\sigma_G^2/\overline{G}^2$ was a free parameter. However, this
parametrization does not describe the data for wide enough range
of {\it{N}}. In further study the linear parametrization of the
COG factor was used. This parametrization was validated on
reasonable interval of {\it{N}}.

\section{Center-of-gravity error parametrization}

Detected position of charged particle  is a random variable given
by several stochastic processes: diffusion, angular effect, gas
gain fluctuation, Landau fluctuation of the secondary ionization,
$\mathbf{E{\times}B}$ effect, electronic noise and systematic
effects (like space charge, etc.). The relative influence of these
processes to the resulting distortion of position determination
depends on the detector parameters. In the big drift detectors
like the ALICE TPC the main contribution is given by diffusion,
gas gain fluctuation, angular effect and secondary ionization
fluctuation.

Furthermore we will use following  assumptions:
\begin{itemize}
\item $N_{\rm{prim}}$ primary electrons  are produced at a random
positions $x_i$ along the particle trajectory. \item $n_i-1$
electrons are produced in the process of secondary ionization.
\item Displacement of produced electrons due to the thermalization
is neglected.
\end{itemize}

Each of electrons is characterized by a random vector
$\vec{z}^i_j$
\begin{eqnarray}
    \vec{z}^i_j =\vec{x}^i+\vec{y}^i_j,
\label{eqZtot}
\end{eqnarray}
where {\it{i}} is the index of primary electron cluster and
{\it{j}} is the index of the secondary electron inside of the
primary electron cluster. Random variable $\vec{x}^i$ is a
position where the primary electron was created. The position
$\vec{y}^i_j$ is a random variable specific for each electron.  It
is given mainly by a diffusion.

The center of gravity of the electron  cloud is given:
\begin{eqnarray}
    \lefteqn{\vec{z}_{\rm{COG}}=\frac{1}{\sum_{i=1}^{N_{\rm{prim}}}
    \sum_{j=1}^{n_i}{g_j^i}}
    \sum_{i=1}^{N_{\rm{prim}}}\sum_{j=1}^{n_i}{g_j^i\vec{z}_j^i}=}
    \nonumber\\
    &&{}\frac{1}{\sum_{i=1}^{N_{\rm{prim}}}
    \sum_{j=1}^{n_i}{g_j^i}}
    \sum_{i=1}^{N_{\rm{prim}}}\vec{x}^i\sum_{j=1}^{n_i}{g_j^i}+\nonumber\\
    &&{}\frac{1}{\sum_{i=1}^{N_{\rm{prim}}}
    \sum_{j=1}^{n_i}{g_j^i}}
    \sum_{i=1}^{N_{\rm{prim}}}\sum_{j=1}^{n_i}{g_j^i\vec{y}_j^i}=
    \nonumber\\ \nonumber\\
    &&{}
    \vec{x}_{\rm{COG}}+\vec{y}_{\rm{COG}}.
\label{eqCOGSec}
\end{eqnarray}

The mean value $\overline{\vec{z}_{\rm{COG}}}$ is equal to the sum
of mean values $\overline{\vec{x}_{\rm{COG}}}$ and
$\overline{\vec{y}_{\rm{COG}}}$.

The sigma of COG in one of the dimension of vector
$\vec{z}_{1COG}$ is given by following equation
\begin{eqnarray}
    \lefteqn{\sigma_{z_{\rm{1COG}}}^2=\sigma_{x_{\rm{1COG}}}^2+
    \sigma_{y_{\rm{1COG}}}^2+}\nonumber\\
    &&{}
        2\left(\overline{x_{\rm{1COG}}y_{\rm{1COG}}}-\bar{x}_{\rm{1COG}}
        \bar{y}_{1COG}\right).
\label{eqCOGSigSec}
\end{eqnarray}

If the vectors $\vec{x}$ and $\vec{y}$ are independent random
variables the last term in the equation (\ref{eqCOGSigSec}) is
equal to zero.
\begin{eqnarray}
    \sigma_{z_{1COG}}^2=\sigma_{x_{\rm{1COG}}}^2+
    \sigma_{y_{\rm{1COG}}}^2,
\label{eqCOGSigSecIn}
\end{eqnarray}
r.m.s. of COG distribution is given by the sum of r.m.s of
{\it{x}} and {\it{y}} components.

In order to estimate the influence of the $\mathbf{E{\times}B}$
and unisochronity effect to the space resolution  two additional
random vectors are added to the initial electron position.

\begin{eqnarray}
\vec{z}^i_j =\vec{x}^i+\vec{y}^i_j+
        \vec{X}_{\mathbf{E{\times}B}}(\vec{x}^i+\vec{y}^i_j)+
        \vec{X}_{\rm{Unisochron}}(\vec{x}^i+\vec{y}^i_j).
\label{eqZtotplus}
\end{eqnarray}
The probability distributions of $\vec{X}_{\mathbf{E{\times}B}}$
and $\vec{X}_{\rm{Unisochron}}$ are  functions of  random vectors
$\vec{x^i}$ and $\vec{y^i_j}$, and they are strongly correlated.
However, simulation indicates that in large drift detectors
distortions, due to these effects,  are negligible compared with a
previous one.

Combining previous equation and neglecting $\mathbf{E{\times}B}$
and unisochronity
effects, the COG distortion  parametrization appears as:\\
{$\sigma_{z}$} of cluster center in {\it{z}} (time) direction
\begin{eqnarray}\
     \lefteqn{\sigma^2_{{z_{\rm{COG}}}} = \frac{D^2_{\rm{L}}
     L_{\rm{Drift}}}{N_{\rm{ch}}}G_{\rm{g}}+}\nonumber\\&&{}
        \frac{{\tan^2\alpha}~L_{\rm{pad}}^2G_{\rm{Lfactor}}(N_{\rm{chprim}})}{12N_{\rm{chprim}}}+
        \sigma^2_{\rm{noise}},
         \label{eqResZ1}
\end{eqnarray}

and {$\sigma_{y}$} of cluster center in {\it{y}}(pad) direction
    \begin{eqnarray}
     \lefteqn{\sigma^2_{y_{\rm{COG}}} = \frac{D^2_{\rm{T}}L_{\rm{Drift}}}{N_{\rm{ch}}}G_{\rm{g}}+}\nonumber\\&&{}
        \frac{{\tan^2\beta}~L_{\rm{pad}}^2G_{\rm{Lfactor}}(N_{\rm{chprim}})}{12N_{\rm{chprim}}}+
        \sigma^2_{\rm{noise}},
        \label{eqResY1}
    \end{eqnarray}
 where
${N_{\rm{ch}}}$ is the total number of electrons in the cluster,
${N_{\rm{chprim}}}$ is the number of primary electrons in the
cluster, ${G_{\rm{g}}}$ is the gas gain fluctuation factor,
${G_{\rm{Lfactor}}}$ is the secondary ionization fluctuation
factor and $\sigma_{\rm{noise}}$ describe the contribution of the
electronic noise to the resulting sigma of the COG.

\section{Precision of cluster COG determination using measured
amplitude}

We have derived parametrization using as parameters the total
number of electrons ${N_{\rm{ch}}}$ and the number of primary
electrons ${N_{\rm{chprim}}}$. This parametrization is in good
agreement with simulated data, where the ${N_{\rm{ch}}}$ and
${N_{\rm{chprim}}}$ are known. It can be  used as an estimate for
the limits of accuracy, if the mean values
$\overline{N}_{\rm{ch}}$ and $\overline{N}_{\rm{chprim}}$ are used
instead.

The ${N_{\rm{ch}}}$ and ${N_{\rm{chprim}}}$ are random variables
described by a Landau distribution, and  Poisson distribution
respectively .

In order to use previously derived formulas (\ref{eqResZ1},
\ref{eqResY1}), the number of electrons can be estimated  assuming
their proportionality to the total measured charge $A$ in the
cluster. However, it turns out that an empirical parametrization
of the factors $G(N)/N=G(A)/(kA)$ gives better results.
Formulas (\ref{eqResZ1}) and (\ref{eqResY1}) are transformed to following form:\\

{$\sigma_{z}$} of cluster center in {\it{z}} (time) direction:
    \begin{eqnarray}
     \lefteqn{\sigma^2_{z_{\rm{COG}}} =
     \frac{D^2_{\rm{L}}L_{\rm{Drift}}}{A}{\times}\frac{G_g(A)}{k_{\rm{ch}}}+}\nonumber\\
        &&{}
        \frac{\tan^2\alpha~L_{\rm{pad}}^2}{12A}{\times}\frac{G_{Lfactor}(A)}{k_{\rm{prim}}}+\sigma^2_{\rm{noise}}
        \label{eqZtotAmp}
    \end{eqnarray}

and {$\sigma_{y}$} of cluster center in {\it{y}}(pad) direction:
    \begin{eqnarray}
     \lefteqn{\sigma_{y_{\rm{COG}}} =
     \frac{D^2_{\rm{T}}L_{\rm{Drift}}}{A}{\times}\frac{G_g(A)}{k_{\rm{ch}}}+}\nonumber\\
        &&{}
        \frac{\tan^2\beta~L_{\rm{pad}}^2}{12A}{\times}\frac{G_{Lfactor}(A)}{k_{\rm{prim}}}+\sigma^2_{\rm{noise}}
        \label{eqYtotAmp}
    \end{eqnarray}

\section{Estimation of the precision of cluster  position
determination using measured cluster shape}

The shape of the cluster is given by the convolution of the
responses to the electron avalanches. The time response function
and the pad response function are almost gaussian, as well as the
spread of electrons due to the diffusion. The spread due to the
angular effect is uniform. Assuming that the contribution of the
angular spread does not dominate the cluster width, the cluster
shape is not far from gaussian. Therefore, we can use the
parametrization

\begin{equation}
       f(t,p) = K_{\rm{Max}}.\exp\left(-\frac{(t-t_{\rm{0}})^2}{2\sigma_{\rm{t}}^2}-
            \frac{(p-p_{\rm{0}})^2}{2\sigma_{\rm{p}}^2}\right),
            \label{eq:GaussTP}
\end{equation}
where  ${K_{\rm{Max}}}$ is the  normalization factor, $t$ and $p$
are time and pad bins, $t_0$ and $p_0$ are centers of the cluster
in time and pad direction and $\sigma_{\rm{t}}$ and
$\sigma_{\rm{p}}$ are the r.m.s. of the time and pad cluster
distribution.

 The mean width of the cluster distribution is given by:
\begin{equation}
     \sigma_{\rm{t}} = \sqrt{D{\rm{^2_L}}L_{\rm{drift}}+\sigma^2_{\rm{preamp}}+
        \frac{\tan^2\alpha~L_{\rm{pad}}^2}{12}},
\end{equation}

\begin{equation}
     \sigma_{\rm{p}} = \sqrt{D{\rm{^2_T}}L_{\rm{drift}}+\sigma^2_{\rm{PRF}}+
        \frac{\tan^2\beta~L_{\rm{pad}}^2}{12}},
\end{equation}
where ${\sigma_{\rm{preamp}}}$ and ${\sigma_{\rm{PRF}}}$  are the
r.m.s. of the time response function and  pad response function,
respectively.

The fluctuation of the shape depends on the contribution of the
random diffusion and angular spread, and on the contribution given
by a gas gain fluctuation and secondary ionization. The
fluctuation of the time and pad response functions is small
compared with the previous one.

The measured r.m.s of the cluster is influenced by a threshold
effect.
\begin{equation}
     \sigma_{\rm{t}}^2 = \sum_{A(t,p)>\rm{threshold}}{(t-t_{\rm{0}})^2{\times}A(t,p)}
\end{equation}
The threshold effect can be eliminated using two dimensional
gaussian fit instead of the simple COG method. However, this
approach is slow and, moreover, the result is very sensitive to
the gain fluctuation.

To eliminate the threshold effect in r.m.s. method, the bins
bellow threshold are replaced with a virtual charge  using
gaussian interpolation of the cluster shape. The introduction of
the virtual charge improves the precision of the COG measurement.
Large systematic shifts in the estimate of the cluster position
(depending on the local track position relative to pad--time) due
to the threshold are no longer observed.

Measuring the r.m.s. of the cluster, the local diffusion and
angular spread of the electron cloud can be estimated. This
provides  additional information for the estimation of
distortions. A simple additional correction function is used:
\begin{eqnarray}
     \sigma_{\rm{COG}} \rightarrow
     \sigma_{\rm{COG}}(A){\times}(1+{\rm{const} {\times}\frac{\delta
     \rm{RMS}}{\rm{teorRMS}}}),
\label{eqResUsingRMS}
\end{eqnarray}
where $\sigma_{\rm{COG}}(A)$ is calculated according formulas
\ref{eqResY1} and \ref{eqResZ1}, and the
$\delta\rm{RMS}/\rm{teorRMS}$ is the relative distortion of the
signal shape from the expected one.

\section{TPC cluster finder}

The classical approach for the beginning of the tracking was
chosen. Before the tracking itself, two-dimensional clusters in
pad-row--time planes are found. Then the positions of the
corresponding space points are reconstructed, which are
interpreted as the crossing points of the tracks and the centers
of the pad rows. We investigate the region 5$\times$5 bins in
pad-row--time plane around the central bin with maximum amplitude.
The size of region, 5$\times$5 bins, is bigger than typical size
of cluster as the $\sigma_{\rm{t}}$ and $\sigma_{\rm{pad}}$ are
about 0.75 bins.

The COG and r.m.s are used to characterize cluster. The COG and
r.m.s are affected by systematic distortions induced by the
threshold effect. Depending on the number of time bins and pads in
clusters the COG and r.m.s. are affected in different ways.
Unfortunately, the number of bins in cluster is the function of
local track position. To get rid of this effect, two-dimensional
gaussian fitting can be used.

Similar results can be achieved  by so called r.m.s. fitting using
virtual charge. The signal below threshold is replaced by the
virtual charge, its expected value according a interpolation. If
the virtual charge is above the threshold value, then it is
replaced with amplitude equal to the threshold value. The signal
r.m.s is used for later error estimation  and as a criteria for
cluster unfolding. This method gives comparable results as
gaussian fit of the cluster but is much faster. Moreover, the COG
position is less sensitive to the gain fluctuations.

The cluster shape depends on the track parameters.  The response
function contribution and diffusion contribution to the cluster
r.m.s. are known during clustering. This is not true for a angular
contribution to the cluster width. The cluster finder should be
optimised for high momentum particle coming from the primary
vertex. Therefore, a conservative approach was chosen, assuming
angle $\alpha$ to be zero. The tangent of the angle $\beta$ is
given by  {\it{z}}-position and pad-row radius, which is known
during clustering.

\subsection{Cluster unfolding}

The estimated width of the cluster is used as criteria for cluster
unfolding. If the r.m.s. in one of the directions is greater then
critical r.m.s,  cluster is considered for unfolding. The fast
spline method is used here. We require the charge to be conserved
in this method. Overlapped clusters  are supposed to have the same
r.m.s., which is equivalent to the same track angles. If this
assumption is not fulfilled, tracks diverge very rapidly.

\begin{figure}[t]
\centering
\includegraphics[width=60mm,angle=-90]{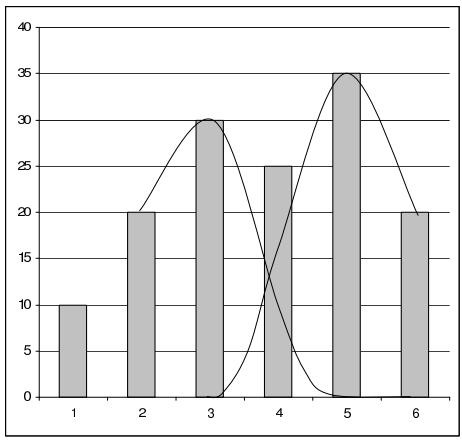}
\caption{
Schematic view of unfolding principle.} \label{figUnfolding1}
\end{figure}
\begin{figure}[t]
\centering
\includegraphics[width=60mm,angle=-90]{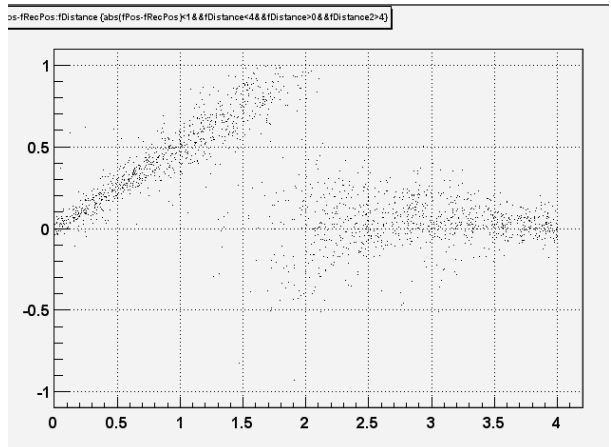}
\caption{ Dependence of the position residual as function of the
distance to the second cluster.} \label{figUnfoldingRes}
\end{figure}

The unfolding algorithm has the following steps:
\begin{itemize}

\item Six amplitudes $C_i$ are investigated (see fig.
\ref{figUnfolding1}). First (left) local maxima, corresponding to
the first cluster is placed at position 3, second (right) local
maxima corresponding to the second cluster is at position 5.

\item In the first iteration, amplitude in bin 4 corresponding to
the cluster on left side $A_{\rm{L4}}$ is calculated using
polynomial interpolation, assuming virtual amplitude at
$A_{\rm{L5}}$ and derivation at $A_{\rm{L5}}^{'}$ to be 0.
Amplitudes $A_{\rm{L2}}$ and $A_{\rm{L3}}$ are considered to be
not influenced by overlap ($A_{\rm{L2}}=C_2$ and
$A_{\rm{L3}}=C_3)$.

\item The amplitude $A_{\rm{R4}}$ is calculated in similar way. In
the next iteration the amplitude $A_{\rm{L4}}$ is calculated
requiring charge conservation
$C_{\rm{4}}=A_{\rm{R4}}+A_{\rm{L4}}$. Consequently
\begin{eqnarray}
   A_{\rm{L4}} \rightarrow
   C_{\rm{4}}\frac{A_{\rm{L4}}}{A_{\rm{L4}}+A_{\rm{R4}}}
\end{eqnarray}
and
\begin{eqnarray}
   A_{\rm{R4}} \rightarrow
   C_{\rm{4}}\frac{A_{\rm{R4}}}{A_{\rm{L4}}+A_{\rm{R4}}}.
\end{eqnarray}
\end{itemize}

Two cluster resolution depends on the distance between the two
tracks. Until  the shape of cluster triggers unfolding, there is a
systematic shifts towards to the COG of two tracks (see fig.
\ref{figUnfoldingRes}), only one cluster is reconstructed.
Afterwards, no systematic shift is observed.

\subsection{Cluster characteristics}

The cluster is characterized by the COG in {\it{y}} and {\it{z}}
directions (fY and fZ) and  by the cluster width (fSigmaY,
fSigmaZ). The deposited charge is described by the signal at
maximum (fMax), and total charge in cluster (fQ). The cluster type
is characterized by the data member fCType which is defined as a
ratio of the charge supposed to be deposited by the track and
total charge in cluster in investigated region 5$\times$5. The
error of the cluster position is assigned to the cluster only
during tracking according formulas
 (\ref{eqZtotAmp}) and  (\ref{eqYtotAmp}), when track
 angles $\alpha$ and $\beta$ are known with sufficient precision.

Obviously, measuring the position of each electron separately the
effect of the gas gain fluctuation can be removed, however this is
not easy to implement in the large TPC detectors. Additional
information about cluster asymmetry can be used, but the resulting
improvement of around 5\% in precision on simulated data  is
negligible, and it is questionable, how successful will be such
correction for the cluster asymmetry on real data.

However, a cluster asymmetry can be used as additional  criteria
for cluster unfolding. Let's denote $\mu_i$ the {\it{i}}-th
central momentum of the cluster, which was created by overlapping
from two sub-clusters with unknown positions and deposited energy
(with momenta $^1\mu_i$ and $^2\mu_i$).

Let $r_1$ is the ratio of two clusters amplitudes:
\[r_1={^1\mu_0}/({^1\mu_0}+{^2\mu_0})\] and the track  distance {\it{d}} is equal to
\[d = {^1\mu_1} -{^2\mu_1}.\]

Assuming that the second moments for both sub-clusters are the
same (${^0\mu_2}={^1\mu_2}={^2\mu_2}$), two sub-clusters distance
{\it{d}} and amplitude ratio $r_1$ can be estimated:
\begin{eqnarray}
     R   = \frac{(\mu_3^6)}{(\mu_2^2-{^0\mu_2^2})^3}\\
    r_{\rm{1}} =0.5\pm0.5{\times}\sqrt{\frac{1}{1-4/R}}  \\
    d   = \sqrt{(4+R){\times}(\mu_2^2-{^0\mu_2^2})}
\label{eqMeas}
\end{eqnarray}

In order to trigger unfolding using the shape information
additional  information about track and mean cluster shape over
several pad-rows are needed. This information is available only
during tracking procedure.

\subsection{TPC seed finding}

The first and the most time-consuming step in tracking is seed
finding. Two different seeding strategies are used, combinatorial
seeding with vertex constraint and simple track follower.

\subsection{Combinatorial seeding algorithm}

Combinatorial seeding starts with a search for all pairs of points
in the pad-row number {\it{i}}1 and in a pad-row {\it{i}}2, $n$
rows closer to the interaction point ($n=i1-i2=20$ at present)
which can project to the primary vertex. The position of the
primary vertex is reconstructed, with high precision, from hits in
the ITS pixel layers, independently of the track determination in
the TPC.

\begin{figure}[t]
\includegraphics[width=60mm]{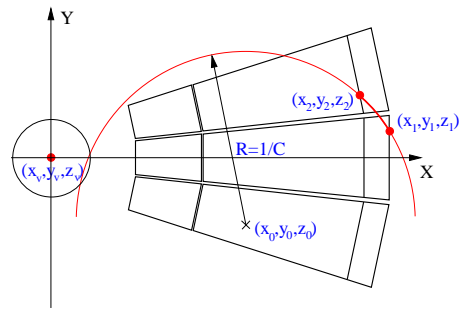}
\caption{ Schematic view of the combinatorial seeding procedure}
\label{figCombSeed}
\end{figure}

Algorithm of combinatorial seeding consists of following steps;

\begin{itemize}
\item Loop over all clusters on pad-row {\it{i}}1
\begin{itemize}
\item Loop over all clusters on pad-row {\it{i}}2, inside a given
window. The size of the window is defined by a cut on track
curvature ({\it{C}}), requiring to seed primary tracks with
$p_{\rm{t}}$ above a threshold.

\begin{itemize}
\item When a reasonable pair of clusters is found, parameters of a
helix going through these points and the primary vertex are
calculated. Parameters of this helix are taken as an initial
approximation of the parameters of the potential track. The
corresponding covariance matrix is evaluated using  the point
errors, which are given by the cluster finder, and applying an
uncertainty of the primary vertex position.  This is the only
place where a certain (not too strong) vertex constraint was
introduced. Later on, tracks are allowed to have any impact
parameters at primary vertex in both the $z$-direction and in
$r$-$\varphi$ plane.

\item Using the calculated helix parameters and their covariance
matrix the Kalman filter is started from the outer point of the
pair to the inner one.

\item If at least half of the potential points between the initial
ones were successfully associated with the track candidate, the
track is saved as a seed.
\end{itemize}
\item End of loop over pad-row 2
\end{itemize}
\item End of loop over pad-row 1
\end{itemize}

\subsection{Track following seeding algorithm} Seeding between
two pad-rows, {\it{i}}1 and {\it{i}}2, starts in the middle
pad-row. For each cluster in the middle pad-row, the two nearest
clusters in the pad-row up and down are found. Afterwards, a
linear fit in both directions ({\it{z}} and {\it{y}}) is
calculated. Expected prolongation to the next two pad-rows are
calculated. For next prolongation again two nearest clusters are
found. Algorithm continue recursively up to the pad-rows {\it{i}}1
and {\it{i}}2. The linear fit is replaced by polynomial after 7
clusters. If more than half of the potential clusters are found,
the track parameters and covariance are calculated as before.

\subsection{Seed finding strategy}

\begin{table}[h]
\centering \caption{Combinatorial seeding efficiency and time
consumption as a function of the  distance between two pad-rows.}
\vglue0.2cm
\begin{tabular}{|c|c|c|}
\hline distance&time &efficiency[\%]\\
\hline 24&  95s&92.2\\
\hline 20&  52s&90.4\\
\hline 16&  34s&88.7\\
\hline 14&  25s&88.1\\
\hline 12&  19s&85.2\\
\hline
\end{tabular}
\label{ch4_tab:dedx}
\end{table}

The main advantage of combinatorial seeding is high efficiency,
around 90\% for primaries with  $p_{\rm{t}}>200\rm{MeV/c}$. The
main disadvantage is the $N^2$ problem of the combinatorial
search. The $N^2$ problem can be reduced restricting the size of
the seeding window. This should be achieved by making the distance
between seeding pad-rows smaller as the size of the window is
proportional to $i2-i1$. However, decreasing the seeding distance,
efficiency of seeding and also quality of seeds deteriorates. The
size of the window can be reduced also by reducing the threshold
curvature  of the track candidate.

However, vertex constraint suppresses secondaries, which should be
 found also. The track following seeding has to be used for them.
This strategy is much faster  but less efficient (80\%). The
efficiency is decreased mainly due to effect of track overlaps and
for low-$p_{\rm{t}}$ tracks by angular effect, which correlates
the cluster position distortion between neighborhood pad-rows.

The efficiency of seeding can be increased repeating of the
seeding procedure in different layers of the TPC. Assuming that
overlapped tracks are random background for the track which should
be seeded, the total efficiency of the seeding can be expressed as

\[\epsilon_{\rm{all}} = 1 - \prod{(1-\epsilon_i)},\]
where $\epsilon_i$ is a efficiency of one seeding. Repeating
seeding, efficiency should reach up to 100\%. Unfortunately,
tracks are sometimes very close on the long path and seeding in
different layers can not be considered as independent. The
efficiency of seeding saturate at a smaller value then 1. Another
problem with repetitive seeding is that occupancy increases
towards to the lower pad-row radius and thus the efficiency is a
function of a the pad-row radius.

However, in order to find  secondaries from kinks or V0 decay, it
is necessary to make a high efficient seeding in outermost
pad-rows. On the other hand in the case of kinks, in the high
density environment it is almost impossible to start tracking of
the primary particles using only the last point of the secondary
track because this point is not well defined. In order to find
them, seeding in innermost pad-rows should be performed. In both
seeding strategies, large decrease of efficiency and precision due
to the dead zones is observed. Additional seeding at the sector
edges is necessary. The length of the pads for the outermost 30
pad-rows is greater than for the other pad-rows. The minimum of
the occupancy and the maximum of seeding efficiency is obtained
when we use outer pad-rows. In order to maximize tracking
efficiency for secondaries it is necessary to make almost
continual seeding inside of the TPC. Several combination of the
slow combinatorial and the fast seeding were investigated.
Depending on the required efficiency, different amount of the time
for seeding can be spent. The default seeding for tracking
performance results was chosen as following: two combinatorial
seedings at outermost 20 pad-rows, and six track following
seedings homogenously spaced inside the outermost sector.

More sophisticated and  faster seeding is currently under
development. It is planned to use, for seeding, only the clusters
which were not assigned to tracks classified as almost perfect.
The criteria for  the almost perfect track  has to be defined,
depending on track density.

\section{Parallel Kalman tracking}

After  seeding, several track hypothesis are tracked in parallel.
Following algorithm is used:
\begin{itemize}

\item For each track candidate the prolongation to the next
pad-row is found.

\item Find nearest cluster.

\item Estimate the cluster position distortions according track
and cluster parameters.

\item Update track according current cluster parameters and
errors.

\item Remove overlapped track hypotheses, i.e. those which share
too many clusters together.

\item Stop not active hypotheses.

\item Continue down to the last pad-row.
\end{itemize}

The prolongation to the next pad-row is calculated according
current track hypothesis. Distortions of the local track position
$\sigma_y$ and $\sigma_x$  are calculated according covariance
matrix. For each track prolongation a window is calculated. The
width of the window is set to $\pm$4$\sigma$ where $\sigma$ is
given by the convolution of the predicted track error and
predicted expectation for cluster r.m.s. Clusters in the container
are ordered according coordinates, binomial search with
log({\it{n}}) performance is used. The nearest  cluster is taken
maximal probable. No cluster competition is currently implemented
because of the memory required when branching the Kalman track
hypothesis and because of the performance penalty.

The width of the search window was chosen to take into account
also overlapped clusters. The position error in this case could be
significantly larger than estimated error for not overlapped
cluster, and the overlap factor is not known apriori. On the other
hand, the minimal distance between two reconstructed clusters is
restricted by a local maxima requirement. Two clusters with
distance less the $\sim$2~bins ($\sim$1~cm) can not be observed.

Once, the nearest cluster is found the cluster error is estimated
using the cluster position and  the amplitude according formulas
(\ref{eqYtotAmp}) and (\ref{eqZtotAmp}). The correction for the
cluster shape and overlapped factor is calculated according
formula (\ref{eqResUsingRMS}).

The cluster is finally accepted  if the square of residuals in
both direction is smaller than estimated 3$\sigma$. If this is the
case track parameters are updated according cluster position and
the error estimates.

It may occur that the track leaves the TPC sector and enters
another one. In this case the track parameters and the covariance
matrix is recalculated  so that they are always expressed in the
local coordinate system of the sector within which the track is at
that moment. The variable fNFindable is defined as a number of
potentially findable clusters. If track is locally inside the
sensitive volume, the fNFindable is incremented otherwise remains
unchanged.

If there are no clusters found in several pad-rows in active
region of the TPC, track hypothesis should be removed. The cluster
density is defined to measure the density of accepted clusters to
all findable clusters in the region, where region is several
pad-rows.

It is not known apriori, if a given track is primary or secondary,
therefore local density can not be interpreted definitely as real
density. This would be true only for tracks which really go
through all considered pad-rows. Tracks with low local density are
not completely removed, they are only signed (fRemoval variable)
for the next analysis.

In order to be able to remove track hypotheses which are almost
the same so called overlap factor is defined. It is the ratio of
the  clusters shared between two tracks candidates and the number
of all clusters. If the overlap factor is greater than the
threshold, track candidate with higher $\chi$2 or significantly
lower number of points is removed. The threshold is parameter,
currently we use the value (in performance studies) at  0.6. This
is a compromise between the maximal efficiency requirement and
minimal number of double found tracks requirement. In the future
this parameters will be optimized, to increase double track
resolution. In this case a new criteria to remove double found
tracks will have to be used.

\subsection{Double track resolution}

In the ALICE TPC represents the main challenge  for tracking the
large track density. From some distance between two tracks the
clusters are not resolved anymore. In our algorithm the track
candidates are removed if some fraction of the clusters are common
to two track candidates.  There are three possibilities, if the
two tracks are overlapped on a very long path. Either it is the
same track, or the two very close tracks or the two tracks where
one changed direction to the second one, and the change of the
direction was misinterpreted as multiple scattering.

New criteria should be defined to handle this situation. Cluster
shape can be used again for this purpose. If the two tracks
overlap and their separation is too small,  only one cluster is
reconstructed, however, its width  is systematically greater.
Moreover, the  charge deposited in the cluster is also
systematically higher.

Another problem is with double found clusters  mainly at the
low-$p_{\rm{t}}$ region. There are two reasons:
\begin{itemize}
\item The non gaussian tail of Coulomb scattering could change the
direction of the track, track can be lost and found again  during
the next seeding.

 \item Because of large inclination and Landau
fluctuations clusters with double local maxima could be created.
\end{itemize}

In order to maximize double-track resolution, and to minimize the
number of double found tracks, the new criteria (mean local
deposited charge and mean local cluster shape)  are under
investigation.

\subsection{{\rm{d}$E$/\rm{d}$x$} measurement}

To estimate particle mean ionization energy loss d$E$/d$x$,
logarithmic truncated mean is used. Using the current cluster
finder the truncation at 60\% gives the best d$E$/d$x$ resolution.
Currently the amplitudes  at local cluster maxima are used,
instead of the total cluster charge, in order to avoid the
distortion due to the track overlaps. Shared clusters are not used
for the estimate of the d$E$/d$x$ at all.

The measured amplitude is normalized to the track length, given by
angles $\alpha$ and $\beta$ and by the pad length. Specific
normalization factors are used for each pad type as the electronic
parameters (gas gain, pad response function) are different in
different parts of the TPC. The normalization condition requires
the same d$E$/d$x$ inside  each part of the TPC for one track.

Correlation between the measured d$E$/d$x$ and particle
multiplicity was observed. The additional correction function for
the cluster shape was successfully introduced, to take into
account local clusters overlaps.

\begin{table}[t]
%\begin{center}
\centering\begin{tabular}{|l|c|}
\hline \textbf{} & \textbf{no} \\
\hline
$\sigma_{\phi}$[mrad]&1.399$\pm$0.030\\
\hline
$\sigma_{\Theta}$[mrad]&0.997$\pm$0.018\\
\hline
$\sigma_{p_{\rm{t}}}$[\%]&0.881$\pm$0.011\\
\hline
$\sigma_{dEdx}/dEdx$[\%]&6.00$\pm$0.2\\
\hline $\epsilon$ [\%]  &99.0\\
\hline
\end{tabular}
\caption{ \label{tab:TrackerPerform}TPC tracking performance
(dN/dy=4000 charged primaries) }
%\end{center}
\end{table}

\section{Conclusions}
We have described current development in the ALICE TPC tracking
which is one of the most challenging task in this experiment. The
track finding efficiency increases, compared to the previous
attempts, for primary tracks by about 10\%, and even more for
secondary tracks. The main improvement is a consequence of the
sophisticated cluster finding and deconvolution which is based on
detail understanding of the physical processes in the TPC and the
optimal usage of achievable information. Another factor which
helped in efficiency increase, especially for secondary tracks, is
the new seeding procedure. The ALICE TPC tracker fulfil,  and even
exceeds the basic requirement. Further development will be
concentrated on secondary vertexing inside TPC and possible use of
information from other detectors.

\bibliographystyle{unsrt}
\end{document}